%% file: f_number_ius.tex
\documentclass[conference]{IEEEtran}

\usepackage{cite}
\usepackage{amsmath,amssymb,amsfonts}
\usepackage{algorithmic}
\usepackage{graphicx}
\usepackage{textcomp}
\usepackage{xcolor}
\def\BibTeX{{\rm B\kern-.05em{\sc i\kern-.025em b}\kern-.08em
    T\kern-.1667em\lower.7ex\hbox{E}\kern-.125emX}}

\usepackage[T1]{fontenc}
\usepackage[utf8]{inputenc}

\usepackage{scalerel}

\usepackage{siunitx}
\usepackage{tikz}
\usepackage{pgfplots}
\pgfplotsset{compat=newest}
\usetikzlibrary{calc,svg.path,pgfplots.external,arrows.meta,backgrounds}
\usepgfplotslibrary{statistics}

\usepackage{array}	
\usepackage{booktabs}	
\usepackage{dcolumn}	
\usepackage{multirow}
\usepackage{multicol}

\usepackage{marginnote}

\newcolumntype{H}{>{\scriptsize}c}

\usepackage[labelformat=simple,labelfont=bf,font=footnotesize]{subcaption}	

\DeclareCaptionLabelSeparator{periodspace}{.\quad}
\definecolor{orcidlogocol}{HTML}{A6CE39}
\tikzset{
  orcidlogo/.pic={
    \fill[orcidlogocol] svg{M256,128c0,70.7-57.3,128-128,128C57.3,256,0,198.7,0,128C0,57.3,57.3,0,128,0C198.7,0,256,57.3,256,128z};
    \fill[white] svg{M86.3,186.2H70.9V79.1h15.4v48.4V186.2z}
                 svg{M108.9,79.1h41.6c39.6,0,57,28.3,57,53.6c0,27.5-21.5,53.6-56.8,53.6h-41.8V79.1z M124.3,172.4h24.5c34.9,0,42.9-26.5,42.9-39.7c0-21.5-13.7-39.7-43.7-39.7h-23.7V172.4z}
                 svg{M88.7,56.8c0,5.5-4.5,10.1-10.1,10.1c-5.6,0-10.1-4.6-10.1-10.1c0-5.6,4.5-10.1,10.1-10.1C84.2,46.7,88.7,51.3,88.7,56.8z};
  }
}

\newcommand\orcidlink[1]{\href{https://orcid.org/#1}{\mbox{\scalerel*{
\begin{tikzpicture}[yscale=-1,transform shape]
\pic{orcidlogo};
\end{tikzpicture}
}{|}}}}

\usepackage{hyperref}

\hypersetup{%
  pdfauthor={M. F. Schiffner and G. Schmitz},%
  pdftitle={Frequency-Dependent F-Number Increases the Contrast and the Spatial Resolution in Fast Pulse-Echo Ultrasound Imaging},%
  pdfsubject={Lehrstuhl für Medizintechnik, Ruhr-Universität Bochum},%
  pdfkeywords={%
    dynamic aperture, %
    F-number, %
    fast ultrasound imaging, %
    Fourier-domain beamforming, %
    frequency-dependent apodization, %
    grating lobes %
  },%
  colorlinks=true,%
  linktocpage=true,%
  bookmarksopenlevel=1%
}

\usepackage{breakurl}	

\usepackage{acronym}  

\usepackage{cleveref}	  			

\crefname{chapter}{Chapter}{Chapters}		
\Crefname{chapter}{Chapter}{Chapters}		
\crefname{appendix}{Appendix}{Appendices}
\Crefname{appendix}{Appendix}{Appendices}
\crefname{section}{Sect.}{Sections}
\Crefname{section}{Section}{Sections}
\crefname{subsection}{Subsect.}{Subsections}
\Crefname{subsection}{Subsection}{Subsections}
\crefname{subsubsection}{Subsect.}{Subsections}
\Crefname{subsubsection}{Subsection}{Subsections}

\crefname{figure}{Fig.}{Figs.}
\Crefname{figure}{Figure}{Figures}

\crefname{table}{Table}{Tables}
\Crefname{table}{Table}{Tables}

\newcommand{\name}[1]{#1}

\definecolor{RUBBLUE}{cmyk}{1.000,0.500,0.000,0.600}	
\definecolor{RUBGREEN}{cmyk}{0.500,0.000,1.000,0.000}	
\definecolor{RUBGRAY}{cmyk}{0.030,0.030,0.030,0.100}	

\definecolor{RUBBLUE_RGB}{rgb}{0.000,0.208,0.377}
\definecolor{RUBGREEN_RGB}{rgb}{0.553,0.682,0.063}
\definecolor{RUBGRAY_RGB}{rgb}{0.906,0.906,0.906}
\definecolor{RUBGRAYDARK_RGB}{rgb}{0.588,0.588,0.588}

\input{0_acronyms.tex}
\acresetall

\begin{document}

\input{copyright.tex}

\title{Frequency-Dependent F-Number Increases the Contrast and the Spatial Resolution in Fast Pulse-Echo Ultrasound Imaging}

\author{%
  \IEEEauthorblockN{Martin F. Schiffner\,\orcidlink{0000-0002-4896-2757}}%
  \IEEEauthorblockA{%
    \textit{Chair of Medical Engineering}\\
    \textit{Ruhr-University Bochum}\\
    Bochum, Germany\\
    martin.schiffner@rub.de%
  }
  \and
  \IEEEauthorblockN{Georg Schmitz\,\orcidlink{0000-0001-5876-7202}}
  \IEEEauthorblockA{%
    \textit{Chair of Medical Engineering}\\
    \textit{Ruhr-University Bochum}\\
    Bochum, Germany\\
    georg.schmitz@rub.de%
  }
}

\maketitle

\begin{abstract}
  \input{abstract.tex}
\end{abstract}
\acresetall

\begin{IEEEkeywords}
  dynamic aperture,
  $F$-number,
  fast ultrasound imaging,
  Fourier-domain beamforming,
  frequency-dependent apodization,
  grating lobes
\end{IEEEkeywords}

\section{Introduction}
\input{introduction/introduction.tex}

\section{Theory}
\input{theory/theory.tex}

\section{Implementation}
\label{sec:implementation}
\input{implementation/implementation.tex}

\section{Experimental Validation}
\input{experimental_validation/experimental_validation.tex}

\section{Results}
\input{results/results.tex}

\section{Conclusion}
\input{conclusion/conclusion.tex}

\IEEEtriggeratref{3}
\IEEEtriggercmd{\enlargethispage{-160mm}}

\bibliographystyle{IEEEtran}


\end{document}

%% file: 0_acronyms.tex
\acrodef{AAPM}{American Association of Physicists in Medicine}		
\acrodef{ACB}{asyn\-chro\-nous compressed beamformer}			
\acrodef{ACA}{adaptive cross approximation}				
\acrodef{ACR}{American College of Radiology}				
\acrodef{ADC}{analog-to-digital conversion}				
\acrodef{AIC}{analog-to-information conversion}				
\acrodef{BIBO}{bounded-input, bounded-output}				
\acrodef{BIM}{\name{Born} iterative method}				
\acrodef{BP}{basis pursuit}						
\acrodef{BPDN}{basis pursuit denoise}					

\acrodef{BVP}{boundary value problem}					
\acrodef{CAR}{Canadian Association of Radiologists}			
\acrodef{CDF}{cumulative distribution function}				
\acrodef{CGP}{conjugate gradient pursuit}				
\acrodef{CNR}{contrast-to-noise ratio}					
\acrodef{CoSaMP}{compressive sampling matching pursuit}			
\acrodef{CPU}{central processing unit}					
\acrodef{CPV}{\name{Cauchy} principal value}				
\acrodef{CPWC}{coherent plane-wave compounding}				
\acrodef{CS}{compressed sensing}					

\acrodef{CSI}{contrast source inversion}				
\acrodef{CT}{x-ray computed tomography}					
\acrodef{CTTE}[CT-TE]{continuous-time ternary encoding}			
\acrodef{CW}{cylindrical wave}						
\acrodef{DAC}{digital-to-analog conversion}				
\acrodef{DAIC}{diagnostically acceptable irreversible compression}	
\acrodef{DAS}{delay-and-sum}						
\acrodef{DBIM}{distorted \name{Born} iterative method}			
\acrodef{DCT}{discrete cosine transform}				
\acrodef{DICOM}{Digital Imaging and Communications in Medicine}		

\acrodef{DFT}{discrete \name{Fourier} transform}			
\acrodef{DSFT}{discrete space \name{Fourier} transform}			
\acrodef{DWT}{discrete wavelet transform}				
\acrodef{DNA}{deoxyribonucleic acid}					
\acrodef{DRG}{German \name{Röntgen} Society}				
\acrodef{DRIM}{distorted \name{Rytov} iterative method}			
\acrodef{DRS}{digital random sampler}					
\acrodef{ERM}{exploding reflector model}				
\acrodef{ESR}{European Society of Radiology}				
\acrodef{EU}{European Union}						

\acrodef{FBP}{filtered backpropagation}					
\acrodef{FDA}{US Food and Drug Administration}				
\acrodef{FDBF}{frequency domain beamforming}				
\acrodef{FDT}{\name{Fourier} diffraction theorem}			
\acrodef{FEHM}{full extent at half maximum}				
\acrodefplural{FEHM}{full extents at half maximum}			
\acrodef{FFT}{fast \name{Fourier} transform}				
\acrodef{FIR}{finite impulse response}					
\acrodef{FLOP}{floating point operation}				
\acrodef{FMM}{fast multipole method}					
\acrodef{FOCUS}{fast object-oriented C++ ultrasound simulator}		

\acrodef{FOV}{field of view}						
\acrodefplural{FOV}{fields of view}					
\acrodef{FPDE}{fractional partial differential equation}		
\acrodef{FPS}{frames per second}					
\acrodef{FRI}{finite rate of innovation}				
\acrodef{FWHM}{full width at half maximum}				
\acrodefplural{FWHM}{full widths at half maximum}			
\acrodef{FWT}{fast wavelet transform}					
\acrodef{GCNR}[gCNR]{generalized contrast-to-noise ratio}		
\acrodef{GPU}{graphics processing unit}					
\acrodef{GWN}{Gaussian white noise}					
\acrodef{HSV}{hue, saturation, value}					

\acrodef{HTP}{hard thresholding pursuit}				
\acrodef{IHT}{iterative hard thresholding}				
\acrodef{IID}[i.i.d.]{independent and identically distributed}		
\acrodef{IQ}{in-phase and quadrature}					
\acrodef{IRLS}{iteratively reweighted least squares}			
\acrodef{ISP}{inverse scattering problem}				
\acrodef{ISR}{integrated sidelobe ratio}				
\acrodef{JPEG}{Joint Photographic Experts Group}			
\acrodef{JSM}{joint sparsity model}					
\acrodef{KK}{\name{Kramers}-\name{Kronig}}                              

\acrodef{LASSO}{least absolute shrinkage and selection operator}	
\acrodef{LDT}{\name{Laplace} diffraction theorem}			
\acrodef{LS}{\name{Lippmann}-\name{Schwinger}}				
\acrodef{LSB}{least significant bit}					
\acrodef{LTI}{linear time-shift invariant}				
\acrodef{MAP}{maximum \emph{a posteriori}}				
\acrodef{MCMC}{\name{Markov} chain Monte Carlo}				
\acrodef{MIRF}{material impulse response function}			
\acrodef{MLFMA}{multilevel fast multipole algorithm}			
\acrodef{MMV}{multiple measurement vectors}				

\acrodef{MOM}[MoM]{method of moments}					
\acrodef{MP}{matching pursuit}						
\acrodef{MP3}{MPEG Audio Layer III}					
\acrodef{MRI}{magnetic resonance imaging}				
\acrodef{MTF}{material transfer function}				
\acrodef{MV}{minimum variance}						
\acrodef{MVDR}{minimum-variance distortionless response}		
\acrodef{MWC}{modulated wideband converter}				
\acrodef{NESTA}{\name{Nesterov}'s algorithm}				
\acrodef{NSP}{null space property}					

\acrodef{NNZC}{number of nonzero components}				
\acrodef{OMP}{orthogonal matching pursuit}				
\acrodef{PAM}{pulse-amplitude modulation}				
\acrodef{PAT}{photoacoustic tomography}					
\acrodef{PC}{personal computer}						
\acrodef{PDE}{partial differential equation}				
\acrodef{PDF}{probability density function}				
\acrodef{PET}{positron emission tomography}				
\acrodef{PICMUS}{Plane-Wave Imaging Challenge in Medical Ultrasound}	
\acrodef{PSF}{point spread function}					

\acrodef{PSL}{peak sidelobe level}					
\acrodef{PSNR}{peak signal-to-noise ratio}				
\acrodef{PW}{plane wave}						
\acrodef{QCW}{quasi-cylindrical wave}					
\acrodef{QPW}{quasi-plane wave}						
\acrodef{QSW}{quasi-spherical wave}					
\acrodef{QQ}{quantile-quantile}						
\acrodef{RA}{randomly-apodized}						
\acrodef{RAD}{randomly-apodized and randomly-delayed}			
\acrodef{RADAR}[RADAR]{radio detection and ranging}			

\acrodef{RAM}{random-access memory}					
\acrodef{RCR}{Royal College of Radiologists}				
\acrodef{RD}{randomly-delayed}						
\acrodef{RF}{radio frequency}						
\acrodef{RIC}{restricted isometry constant}				
\acrodef{RIP}{restricted isometry property}				
\acrodef{RMSE}{root mean-squared error}					
\acrodef{ROI}{region of interest}					
\acrodefplural{ROI}{regions of interest}				
\acrodef{ROMP}{regularized \ac{OMP}}					
\acrodef{RS}{\name{Rayleigh}-\name{Sommerfeld}}                         

\acrodef{SaS}[S$\alpha$S]{symmetric $\alpha$-stable}			
\acrodef{SA}{synthetic aperture}					
\acrodef{SBL}{sparse Bayesian learning}					
\acrodef{SDK}{software development kit}					
\acrodef{SDR}{software-defined radio}					
\acrodef{SIIM}{Society for Imaging Informatics in Medicine}		
\acrodef{SISO}{single-input, single-output}				
\acrodef{SINR}{signal-to-interference-plus-noise ratio}			
\acrodef{SMI}{sample matrix inversion}					
\acrodef{SNR}{signal-to-noise ratio}					

\acrodef{SOMP}{simultaneous orthogonal matching pursuit}		
\acrodef{SoS}{sum of sincs}						
\acrodef{SP}{subspace pursuit}						
\acrodef{spark}{sparse rank}						
\acrodef{SPECT}{single photon emission computed tomography}		
\acrodef{SPG}{spectral projected gradient}				
\acrodef{SPGL1}[SPG$\ell_{1}$]{spectral projected gradient for $\ell_{1}$-minimization}	
\acrodef{SPLS}{split-projection least squares}				
\acrodef{SR}{sparse recovery}						
\acrodef{SRC}{\name{Sommerfeld} radiation condition}			

\acrodef{SSI}{supersonic shear imaging}					
\acrodef{SSIM}{structural similarity}					
\acrodef{STFT}{short-time \name{Fourier} transform}			
\acrodef{StOMP}{stagewise \ac{OMP}}					
\acrodef{StWGP}{stagewise weak \ac{CGP}}				
\acrodef{SVD}{singular value decomposition}				
\acrodef{TDMA}{time-division multiple access}				
\acrodef{TMSBL}[T-MSBL]{temporally correlated \ac{MMV}-based \ac{SBL}}  
\acrodef{TOF}{time-of-flight}						
\acrodefplural{TOF}{times-of-flight}					
\acrodef{TGC}{time gain compensation}					

\acrodef{THI}{tissue harmonic imaging}					
\acrodef{TPSF}{transform point spread function}				
\acrodef{TR}{\name{Tikhonov} regularization}				
\acrodef{TV}{total variation}						
\acrodef{UCA}{ultrasound contrast agents}				
\acrodef{UDT}{ultrasound diffraction tomography}			
\acrodef{UI}{ultrasound imaging}					
\acrodef{UoS}{union of subspaces}					
\acrodef{USA}{United States of America}					
\acrodef{UWB}{ultra-wideband}						

\acrodef{Xampling}{CS-Sampling}						

%% file: copyright.tex
\begin{titlepage}
\thispagestyle{empty}%
\noindent
{\huge
Frequency-Dependent F-Number Increases\\
the Contrast and the Spatial Resolution in\\
Fast Pulse-Echo Ultrasound Imaging
}
\par
\vspace{36pt}
\noindent
{\large Martin F. Schiffner and Georg Schmitz\par\vspace{12pt}
\noindent \href{http://www.mt.rub.de}{Chair of Medical Engineering}, Ruhr-University Bochum, 44801 Bochum, Germany}
\vspace{36pt}
\par
\noindent
{\bf Copyright notice:}\par\vspace{12pt}
\noindent
\copyright~2021~IEEE.
Personal use of
this material is
permitted.
Permission from
IEEE must be obtained for
all other uses, in
any current or
future media, including reprinting/republishing
this material for
advertising or
promotional purposes, creating
new collective works, for
resale or
redistribution to
servers or
lists, or
reuse of
any copyrighted component of
this work in
other works.
\par
\vspace{12pt}
\noindent
{\bf Full citation:}\par\vspace{12pt}
\noindent
2021 IEEE Int. Ultrasonics Symp. (IUS), Xi'an, China, Sep. 2021, pp. 1--4.
\par
\noindent
DOI: \href{https://doi.org/10.1109/IUS52206.2021.9593488}{10.1109/IUS52206.2021.9593488}
\par
\vspace{12pt}
\noindent
\href{https://ieeexplore.ieee.org/document/9593488}{Click here for IEEE Xplore}
\clearpage
\end{titlepage}

%% file: abstract.tex
Fixed $F$-numbers reduce
grating lobe artifacts in
fast pulse-echo ultrasound imaging.
Such $F$-numbers result in
dynamic receive subapertures whose widths vary with
the focal position.
These subapertures, however, ignore
useful low-frequency components in
the excluded \ac{RF} signals and, thus, reduce
the lateral resolution.
Here, we propose
a frequency-dependent $F$-number to simultaneously suppress
grating lobe artifacts and maintain
the lateral resolution.
This $F$-number, at
high frequencies, reduces
the receive subaperture to remove
spatially undersampled components of
the \ac{RF} signals and suppress
grating lobes.
The $F$-number, at
low frequencies, enlarges
the receive subaperture to use
the components of
all \ac{RF} signals and maintain
the lateral resolution.
Experiments validated
the proposed $F$-number and demonstrated
improvements in
the contrast and
the widths of
wire targets of up to
\SI{3.2}{\percent} and
\SI{12.8}{\percent},
respectively.

%% file: introduction/introduction.tex
Software-based fast imaging modes combine
high bandwidths with
fully-sampled transducer arrays
\cite{article:TanterITUFFC2014,article:MontaldoITUFFC2009,article:JensenUlt2006}.
This combination, owing to
large element pitch-to-wavelength ratios, suffers from
spatial undersampling and requires
a method to prevent
image degradation.
The formation of receive beams by
small subapertures, which are defined by
a fixed F-number and vary with
the focal position
(see, e.g.,
  \cite{article:PerrotUltrasonics2021},
  \cite[(3)]{article:MontaldoITUFFC2009}%
), avoids
this undersampling and reduces
grating lobe artifacts.
These subapertures, however, ignore
useful low-frequency components in
the excluded \ac{RF} signals and, thus, reduce
the lateral resolution.
Here,
we propose
a frequency-dependent $F$-number to suppress
image artifacts and approximately maintain
the lateral resolution of
the full aperture.
We first derive
a closed-form expression for
this $F$-number from
the far-field directivity pattern of
a focused receive subaperture.
We subsequently outline
a simple Fourier-domain beamforming algorithm
(see, e.g., \cite{%
  software:FNumber,
  proc:SchiffnerIUS2016a%
}) to implement
this $F$-number and show
its benefits in
a phantom experiment.

%% file: theory/theory.tex
%
\begin{figure}[t!]
 \centering%
  \input{theory/figures/latex/theory_f_number.tex}
 \caption{}
 \label{fig:V}
\end{figure}
C
{
 Geometry of
 the uniform linear transducer array and definition of
 the $F$-number
 \eqref{eqn:theory_f_number}.
}%
{theory_f_number}

The following derivations consider
a uniform linear transducer array, as shown in
\cref{fig:theory_f_number}, with
$N_{\text{el}}$ elements of
width $w$ and
the element pitch
$p \geq w$.
A receive subaperture, which focuses
a beam on
the point
$( x_{\text{f}}, z_{\text{f}} )$, consists of
$N_{\text{sub}} \leq N_{\text{el}}$ elements and, unless indicated otherwise, is symmetric about
the lateral focal coordinate
$x_{\text{f}}$.
The usual dispersion relation, i.e.,
$\lambda f = c$, links
the product of
the wavelength
$\lambda$ and
the frequency
$f$ to
the average speed of sound
$c$.

\subsection{What is the $F$-Number?}
\input{theory/theory_f_number.tex}

\subsection{Effect of the $F$-Number on the Image Quality}
\label{subsec:theory_effect_receive_beam}
\input{theory/theory_effect_receive_beam.tex}

\subsection{Frequency Dependence of the Grating Lobe Angles}
\input{theory/theory_far_field_frequency.tex}

\subsection{Proposed $F$-Number}
\input{theory/theory_f_number_proposed.tex}

\subsection{What Does the Proposed $F$-Number Accomplish?}
\input{theory/theory_accomplishments.tex}

%% file: theory/figures/latex/theory_f_number.tex
%
\begin{tikzpicture}%
[%
 node distance = 4.25cm and 4cm,
 font = \footnotesize,
 axes_r/.style = { ->, thick },
 element/.style = { draw = black, thin },
 elementactive/.style = { fill = RUBGREEN_RGB },
 declare function = {
   width_aperture(\n,\p,\w) = (\n - 1) * \p + \w;
   aperture_lb(\n,\p,\w) = - width_aperture(\n,\p,\w) / 2;
   aperture_ub(\n,\p,\w) = + width_aperture(\n,\p,\w) / 2;
 }%
]

\pgfmathsetmacro{\MinLengthX}{2.2}
\pgfmathsetmacro{\MaxLengthX}{2.2}

\pgfmathsetmacro{\AxisYStart}{0.2}
\pgfmathsetmacro{\AxisYStop}{-2.2}

\pgfmathsetmacro{\ElementTickHeight}{0.1}

\pgfmathsetmacro{\NElements}{16}
\pgfmathsetmacro{\ElementPitch}{0.25}
\pgfmathsetmacro{\ElementHeight}{0.4}
\pgfmathsetmacro{\ElementWidth}{0.2}

\pgfmathsetmacro{\FocusX}{1}
\pgfmathsetmacro{\FocusZ}{1.45}

\pgfmathsetmacro{\FNumber}{1}

\pgfmathsetmacro{\WidthElementIndex}{1}
\pgfmathsetmacro{\PitchElementIndex}{3}

\pgfmathsetmacro{\MElements}{ ( \NElements - 1 ) / 2 }
\pgfmathsetmacro{\ApertureWidthOverTwo}{ \FocusZ / ( 2 * \FNumber ) }

\pgfmathsetmacro{\ApertureLB}{ \FocusX - \ApertureWidthOverTwo }
\pgfmathsetmacro{\ApertureUB}{ \FocusX + \ApertureWidthOverTwo }

\tikzset{
  pics/element/.style = {%
    code = {%
      \draw [ element, #1 ] (-\ElementWidth / 2, 0) rectangle (\ElementWidth / 2, \ElementHeight);
    }
  }
}

\newcommand{\fillactive}[1]
{%

  \coordinate (ELB) at ( { #1 - \ElementWidth / 2 }, 0 );
  \coordinate (ERU) at ( { #1 + \ElementWidth / 2 }, \ElementHeight );

  \begin{scope}[on background layer]
    \fill [ elementactive ] (ELB) rectangle (ERU);
  \end{scope}
}

\coordinate (r1AxisMin) at (-\MinLengthX, 0);
\coordinate (r1AxisMax) at (\MaxLengthX, 0);
\coordinate (r2AxisMin) at (0, \AxisYStart);
\coordinate (r2AxisMax) at (0, \AxisYStop);

\draw [ axes_r ] (r1AxisMin) -- (r1AxisMax);	
\draw [ axes_r ] (r2AxisMin) -- (r2AxisMax);	

\foreach \x in {1,2,...,\NElements}{%

  \pgfmathsetmacro{\PosXCenter}{(\x - 1 - \MElements) * \ElementPitch}

  \pic at (\PosXCenter, 0) { element };

  \pgfmathparse{ ( \PosXCenter >= \ApertureLB ) ? 1 : 0 }
  \ifdim\pgfmathresult pt>0pt

  \pgfmathparse{ ( \PosXCenter <= \ApertureUB ) ? 1 : 0 }
  \ifdim\pgfmathresult pt>0pt

    \fillactive{ \PosXCenter }

  \fi 
  \fi 

  \draw [thin] ( \PosXCenter, - \ElementTickHeight / 2 ) -- ( \PosXCenter, \ElementTickHeight / 2 );

} 

\draw [ |-| ] ( { ( \WidthElementIndex - 1 - \MElements ) * \ElementPitch - 0.5 * \ElementWidth }, \ElementHeight + 0.2 )
  -- ( { ( \WidthElementIndex - 1 - \MElements ) * \ElementPitch + 0.5 * \ElementWidth }, \ElementHeight + 0.2 )
  node [ pos = 0.5, anchor = south ] { $w$ };

\draw [ |-| ] ( { ( \PitchElementIndex - 1 - \MElements ) * \ElementPitch }, \ElementHeight + 0.2 )
  -- ( { ( \PitchElementIndex - \MElements ) * \ElementPitch }, \ElementHeight + 0.2 )
  node [ pos = 0.5, anchor = south ] { $p$ };

\draw [ red, <->, >={Latex} ] ( \FocusX, 0 ) -- ( \FocusX, - \FocusZ )
  node [ pos = 0.5, anchor = west ] { $z_{\text{f}}$ };

\draw [ red, <->, >={Latex} ] ( 0, - \FocusZ ) -- ( \FocusX, - \FocusZ )
  node [ pos = 0.5, anchor = north ] { $x_{\text{f}}$ };

\node [ red, font = \normalsize, label = {right:Focus} ] at ( \FocusX, - \FocusZ ) { $+$ };

\draw [ elementactive, RUBGREEN_RGB, |-| ] ( \FocusX - \ApertureWidthOverTwo, \ElementHeight + 0.2 )
  -- ( \FocusX + \ApertureWidthOverTwo, \ElementHeight + 0.2 )
  node [ pos = 0.5, anchor = south ] { $A( z_{\text{f}} )$ };

\node [ right ] at (r1AxisMax) { $x$ };
\node [ below ] at (r2AxisMax) { $z$ };

\end{tikzpicture}

%% file: theory/theory_f_number.tex
The $F$-number, for
a uniform linear transducer array, equals
the quotient of
the focal length
$z_{\text{f}}$ and
the width of
the receive subaperture
$A( z_{\text{f}} )$, i.e.,
\cite{article:PerrotUltrasonics2021},
\cite[(3)]{article:MontaldoITUFFC2009}%
\begin{equation}
  F
  = 
  \frac{
    z_{\text{f}}
  }{
    A( z_{\text{f}} )
  },
 \label{eqn:theory_f_number}
\end{equation}
as shown in
\cref{fig:theory_f_number}.
The usage of
a fixed $F$-number results in
a dynamic receive subaperture whose
width increases with
the focal length.
This increase, however, is limited by
the finite aperture of
the transducer array and, due to
the discrete elements, discontinuous.
The desired $F$-number, for
these reasons, may differ from
the actual $F$-number, especially for
(i)
large focal lengths or
(ii)
lateral focal coordinates
$x_{\text{f}}$ near
the array bounds.
Typical $F$-numbers range from
$1$ to $2$
\cite[p. 414]{book:Szabo2013}.

%% file: theory/theory_effect_receive_beam.tex
%
\begin{figure}[t!]
 \centering%
  \input{theory/figures/latex/theory_beams.tex}
 \caption{}
 \label{fig:V}
\end{figure}
C
{
 Effect of
 the $F$-number
 \eqref{eqn:theory_f_number} on
 a monofrequent receive beam.
 The focal length,
 the element width, and
 the element pitch amount to
 $z_{\text{f}} = 30 \lambda$,
 $w = 0.981 \lambda$, and
 $p = \lambda$,
 respectively.
 The receive subaperture is
 symmetric about
 the lateral focal coordinate
 $x_{\text{f}} = 0$.
}%
{theory_beams}

The $F$-number
\eqref{eqn:theory_f_number} trades off
the lateral resolution against
the suppression of
grating lobe artifacts.
This conclusion may be drawn from
the effect of
the $F$-number on
a monofrequent receive beam, as shown in
\cref{fig:theory_beams}.
The \ac{FWHM} of
the main lobe in
the focal plane decreases with
the $F$-number
(see also, e.g., \cite[(8)]{article:LuUMB1994}).
%
Small $F$-numbers or, equivalently, large receive subapertures thus increase
the lateral resolution.
The first-order grating lobes, as
the $F$-number decreases, however, move toward
the main lobe and, owing to
the directivity of
the array elements, increase in
amplitude.
Large $F$-numbers or, equivalently, small receive subapertures thus reduce
grating lobe artifacts.
%

%% file: theory/figures/latex/theory_beams.tex
%
\begin{tikzpicture}%
[%
 node distance = 4.25cm and 4cm,
 font = \footnotesize,
 axes_base/.style = { -> },
 background/.style = { inner xsep = 1em, inner ysep = 1.2em, thick, draw = #1, draw opacity = 0.2, fill = #1, fill opacity = 0.2, text opacity = 1, rounded corners = 2 },
 matrix_plot/.style = { row sep = 1mm, column sep = 2.5mm },
 focal_plane/.style = { dashed, very thick },
 focus/.style = { red },
 profile_1/.style = { RUBBLUE_RGB },
 profile_2/.style = { RUBGREEN_RGB },
 profile_3/.style = { orange },
 alpha_lb/.style = { red, dashed },
 alpha_ub/.style = { red, dashed },
 alpha_lb_act/.style = { yellow },
 alpha_ub_act/.style = { yellow },
 grating_lobe/.style = { white },
 legend/.style = { draw = black!50, fill = black!5, rounded corners = 5, thick, outer sep = 0 },
 title/.style = { inner sep = 0, outer sep = 0, anchor = base west, font = \footnotesize\sffamily },
 bg_tiles/.style = { fill = black!3, draw = black!30, very thick, rounded corners = 2, inner xsep = 1ex, inner ysep = 1ex },
 aperture_active/.style = { yellow, opacity = 0.6 },
 aperture_inactive/.style = { black!20 },
 title/.style = { inner sep = 0, outer sep = 0, anchor = south west, font = \bfseries },
 declare function = {
   width_aperture(\n,\p,\w) = (\n - 1) * \p + \w;
   aperture_lb(\n,\p,\w) = - width_aperture(\n,\p,\w) / 2;
   aperture_ub(\n,\p,\w) = + width_aperture(\n,\p,\w) / 2;
   fill = 279.8 / 304.8;
   F_number_lb(\r) = sqrt( max( pow( \r, 2 ), 0.25 ) - 0.25 );
   F_number(\z,\A) = \z / \A;
   angular_aperture_lb(\z,\A) = 90 - atan( 1 / ( 2 * F_number(\z,\A) ) );
   angular_aperture_ub(\z,\A) = 90 + atan( 1 / ( 2 * F_number(\z,\A) ) );
   directivity(\r) = fill * \r / 1.2;
   main_bound(\z,\A) = 1 / sqrt( 1 + pow( 2 * F_number(\z,\A), 2 ) );
   grating_lobe_lb(\r,\z,\A) = -1 / \r + 1 / sqrt( 1 + pow( 2 * F_number(\z,\A), 2 ) );
   grating_lobe_ub(\r,\z,\A) = 1 / \r - 1 / sqrt( 1 + pow( 2 * F_number(\z,\A), 2 ) );
 }%
]

\pgfmathsetmacro{\AxisLengthX}{3.2}
\pgfmathsetmacro{\AxisLengthZ}{3.2}

\pgfmathsetmacro{\AxisDistanceX}{1.1}         
\pgfmathsetmacro{\AxisDistanceZ}{1.75}       
\pgfmathsetmacro{\AxisDistanceZColorbar}{1} 
\pgfmathsetmacro{\AxisDistanceZProfile}{0.5} 

\pgfmathsetmacro{\ColorbarLengthY}{0.2}     

\pgfmathsetmacro{\lengthBounds}{1.6}
\pgfmathsetmacro{\legendDistance}{20}
\pgfmathsetmacro{\titleSep}{0.5}

\pgfmathsetmacro{\NElements}{128}
\pgfmathsetmacro{\ElementPitchNorm}{1}
\pgfmathsetmacro{\WidthOverPitch}{0.981}

\pgfmathsetmacro{\ElementHeight}{0.2}

\pgfmathsetmacro{\ImageLimitsXLB}{-100} 
\pgfmathsetmacro{\ImageLimitsXUB}{100}  
\pgfmathsetmacro{\ImageLimitsZLB}{0}  
\pgfmathsetmacro{\ImageLimitsZUB}{200}  

\pgfmathsetmacro{\AxisLimitsXLB}{-100}  
\pgfmathsetmacro{\AxisLimitsXUB}{100}   
\pgfmathsetmacro{\AxisLimitsZLB}{0}     
\pgfmathsetmacro{\AxisLimitsZUB}{120}   

\pgfmathsetmacro{\focalLengthNorm}{30}
\pgfmathsetmacro{\focusLateralLB}{-5}
\pgfmathsetmacro{\focusLateralUB}{5}
\pgfmathsetmacro{\focusAxialLB}{ \focalLengthNorm - 5 }
\pgfmathsetmacro{\focusAxialUB}{ \focalLengthNorm + 5 }

\pgfmathsetmacro{\TitleXShift}{-0.15}   
\pgfmathsetmacro{\TitleZShift}{0.08}     

\pgfmathsetmacro{\AxisLabelXShift}{-0.1}  
\pgfmathsetmacro{\AxisLabelZShift}{-0.15}  

\pgfmathsetmacro{\LegendLengthX}{ 2 * \AxisLengthX + \AxisDistanceX }

\pgfmathsetmacro{\ColorbarLengthX}{ \LegendLengthX }

\pgfmathsetmacro{\AxisWidthPolar}{ \AxisLengthZ - \AxisDistanceZProfile }
\pgfmathsetmacro{\AxisHeightPolar}{\AxisWidthPolar}

\pgfmathsetmacro{\AxisWidthProfile}{ \AxisLengthX }
\pgfmathsetmacro{\AxisHeightProfile}{ \AxisLengthZ }

%

\pgfmathsetmacro{\ElementWidthNorm}{ \WidthOverPitch * \ElementPitchNorm }

\pgfmathsetmacro{\FNumberLB}{ F_number_lb( \ElementPitchNorm ) }

\pgfmathsetmacro{\apertureWidthMax}{ \focalLengthNorm / \FNumberLB }
\pgfmathsetmacro{\apertureXMin}{ -\apertureWidthMax / 2 }
\pgfmathsetmacro{\apertureXMax}{ \apertureWidthMax / 2 }

\pgfmathsetmacro{\alphaLB}{ angular_aperture_lb( \focalLengthNorm, \apertureWidthMax ) }
\pgfmathsetmacro{\alphaUB}{ angular_aperture_ub( \focalLengthNorm, \apertureWidthMax ) }

\pgfmathsetmacro{\FactorScaleLateral}{ \AxisLengthX / ( \AxisLimitsXUB - \AxisLimitsXLB ) }



\tikzset{
  pics/focus/.style n args = {3}{%
    code = {%
      \draw [ focus ] ( -0.1, 0 ) -- ( 0.1, 0 );
      \draw [ focus ] ( 0, -0.1 ) -- ( 0, 0.1 );
    } 
  } 
} 

\tikzset{
  pics/beam_pattern_pic/.style n args = {6}{%
    code = {%
      \begin{axis}
      [
        width = \AxisLengthX cm, height = \AxisLengthZ cm,
        scale only axis = true,
        at = { ( 0, 0 ) },
        anchor = north west,
        xmin = \AxisLimitsXLB, xmax = \AxisLimitsXUB,
        ymin = \AxisLimitsZLB, ymax = \AxisLimitsZUB,
        y dir = reverse,
         axis x line = bottom,
         axis y line = center,
         axis y line shift = -\AxisLimitsXLB,
         axis line style = { - },
         every axis title/.append style = { title, at = { ( \TitleXShift, 1 + \TitleZShift ) } },
         title = { #2 },
         xtick = { -100, -50, 0, 50, 100 },
         xticklabel style = { font = \scriptsize },
         ytick = { 0, 25, 50, 75, 100 },
         yticklabel style = { font = \scriptsize },
        x label style = { at = { (axis description cs:0.5,\AxisLabelXShift) }, anchor = north }, 
        y label style = { at = { (axis description cs:\AxisLabelZShift,0.5) }, anchor = south, rotate = 90 }, 
        xlabel = { #4 },
        ylabel = { #5 },
        name = -axes%
      ]
        \addplot graphics [ xmin = \ImageLimitsXLB,  xmax = \ImageLimitsXUB, ymin = \ImageLimitsZLB, ymax = \ImageLimitsZUB ]
          { #1 };

        \draw [ focal_plane, #6 ] ( \AxisLimitsXLB, \focalLengthNorm ) -- ( \AxisLimitsXUB, \focalLengthNorm );

        \draw [ focus ] (0, \focusAxialLB) -- (0, \focusAxialUB);
        \draw [ focus ] (\focusLateralLB, \focalLengthNorm) -- (\focusLateralUB, \focalLengthNorm);

        \pgfmathsetmacro{\ApertureXLB}{ aperture_lb( #3, \ElementPitchNorm, \ElementWidthNorm ) }
        \pgfmathsetmacro{\ApertureXUB}{ aperture_ub( #3, \ElementPitchNorm, \ElementWidthNorm ) }

        \pgfmathsetmacro{\gratingXLB}{ 100 * grating_lobe_lb( \ElementPitchNorm, \focalLengthNorm, width_aperture( #3, \ElementPitchNorm, \ElementWidthNorm ) ) + \ApertureXLB }
        \pgfmathsetmacro{\gratingZLB}{ 100 * sqrt( 1 - pow( ( \gratingXLB - \ApertureXLB ) / 100, 2 ) ) }
        \draw [white] ( \ApertureXLB, 0 ) -- ( \gratingXLB, \gratingZLB );

        \pgfmathsetmacro{\gratingXUB}{ 100 * grating_lobe_ub( \ElementPitchNorm, \focalLengthNorm, width_aperture( #3, \ElementPitchNorm, \ElementWidthNorm ) ) + \ApertureXUB }
        \pgfmathsetmacro{\gratingZUB}{ 100 * sqrt( 1 - pow( ( \gratingXUB - \ApertureXUB ) / 100, 2 ) ) }
        \draw [white] ( \ApertureXUB, 0 ) -- ( \gratingXUB, \gratingZUB );

        \pgfmathsetmacro{\RadiusAct}{ sqrt( pow( \AxisLimitsZUB - \AxisLimitsZLB, 2 ) + pow( main_bound( \focalLengthNorm, width_aperture( #3, \ElementPitchNorm, \ElementWidthNorm ) ), 2 ) ) }

        \pgfmathsetmacro{\angleLB}{ angular_aperture_lb( \focalLengthNorm, width_aperture( #3, \ElementPitchNorm, \ElementWidthNorm ) ) }
        \pgfmathsetmacro{\stopLBX}{ \RadiusAct * cos( \angleLB ) + \ApertureXLB }
        \pgfmathsetmacro{\stopLBZ}{ \RadiusAct * sin( \angleLB ) }
        \draw [ alpha_lb_act ] (\ApertureXLB,0) -- (\stopLBX,\stopLBZ);

        \pgfmathsetmacro{\angleUB}{ angular_aperture_ub( \focalLengthNorm, width_aperture( #3, \ElementPitchNorm, \ElementWidthNorm ) ) }
        \pgfmathsetmacro{\stopUBX}{ \RadiusAct * cos( \angleUB ) + \ApertureXUB }
        \pgfmathsetmacro{\stopUBZ}{ \RadiusAct * sin( \angleUB ) }
        \draw [ alpha_ub_act ] (\ApertureXUB,0) -- (\stopUBX,\stopUBZ);

%

      \end{axis}

      \pgfmathsetmacro{\NKerfsAll}{ \NElements - 1 }
      \pgfmathsetmacro{\MElementsAll}{ \NKerfsAll / 2 }
      \pgfmathsetmacro{\NKerfs}{ #3 - 1 }
      \pgfmathsetmacro{\MElements}{ \NKerfs / 2 }
      \pgfmathsetmacro{\ArrayXLB}{ aperture_lb( \NElements, \ElementPitchNorm, \ElementWidthNorm ) }
      \pgfmathsetmacro{\ArrayXUB}{ aperture_ub( \NElements, \ElementPitchNorm, \ElementWidthNorm ) }
      \pgfmathsetmacro{\ApertureXLB}{ aperture_lb( #3, \ElementPitchNorm, \ElementWidthNorm ) }
      \pgfmathsetmacro{\ApertureXUB}{ aperture_ub( #3, \ElementPitchNorm, \ElementWidthNorm ) }

      \fill [ aperture_inactive ] ( 0.5 * \AxisLengthX + \FactorScaleLateral * \ArrayXLB, 0 ) rectangle ( 0.5 * \AxisLengthX + \FactorScaleLateral * \ApertureXLB, \ElementHeight );
      \fill [ aperture_active ] ( 0.5 * \AxisLengthX + \FactorScaleLateral * \ApertureXLB, 0 ) rectangle ( 0.5 * \AxisLengthX + \FactorScaleLateral * \ApertureXUB, \ElementHeight );
      \fill [ aperture_inactive ] ( 0.5 * \AxisLengthX + \FactorScaleLateral * \ApertureXUB, 0 ) rectangle ( 0.5 * \AxisLengthX + \FactorScaleLateral * \ArrayXUB, \ElementHeight );

      \foreach \i in {1,...,\NKerfsAll}
      {
        \fill [ black ] ( { 0.5 * \AxisLengthX + \FactorScaleLateral * ( ( \i - 1 - \MElementsAll ) * \ElementPitchNorm + 0.5 * \ElementWidthNorm ) }, 0 ) rectangle ( { 0.5 * \AxisLengthX + \FactorScaleLateral * ( ( \i - \MElementsAll ) * \ElementPitchNorm - 0.5 * \ElementWidthNorm ) }, \ElementHeight );
      }

    } 
  } 
} 

\tikzset{
  pics/profile_pic/.style n args = {7}{%
    code = {%
      \begin{axis}
      [
        width = \AxisWidthProfile cm, height = \AxisHeightProfile cm,
        scale only axis = true,
        at = { ( 0, 0 ) },
        anchor = north west,
        grid = both,
        xmin = \AxisLimitsXLB, xmax = \AxisLimitsXUB,
        ymin = -30, ymax = 0,
        xtick = { -100, -50, 0, 50, 100 },
        ytick = { -30, -25, -20, -15, -10, -5, 0 },
        xticklabel style = { font = \scriptsize },
        yticklabel style = { font = \scriptsize },
        every axis title/.append style = { title, at = { ( \TitleXShift, 1 + \TitleZShift ) } },
        title = { #5 },
        x label style = { at = { (axis description cs:0.5,\AxisLabelXShift) }, anchor = north }, 
        y label style = { at = { (axis description cs:\AxisLabelZShift,0.5) }, anchor = south }, 
        xlabel = { #6 },
        ylabel = { #7 },
        name = -axes%
      ]
        \addplot [ profile_1 ] table[ x = x, y = p, search path = {theory/figures} ]
          {#1};
        \addplot [ profile_2 ] table[ x = x, y = p, search path = {theory/figures} ]
          {#2};
        \addplot [ profile_3 ] table[ x = x, y = p, search path = {theory/figures} ]
          {#3};

      \end{axis}
    } 
  } 
} 

\tikzset{
  pics/colorbar/.style n args = {3}{%
    code = {%
      \begin{axis}
      [
        scale only axis = true,
        width = 0, height = 0,
        at = { ( 0, 0 ) },
        anchor = north west,
        hide axis,
        point meta min = #1, point meta max = #2,
        colormap/blackwhite,
        colorbar horizontal,
        colorbar style = {
          at = { ( 0, 0 ) },
          xtick = { #1, -25, -20, -15, -10, -5, #2 },
          xticklabel style = { font = \scriptsize },
          anchor = north west,
          width = \ColorbarLengthX cm, height = \ColorbarLengthY cm,
          y label style = { at = { (axis description cs:0,0.5) }, anchor = east, rotate = -90 },
          ylabel = { #3 }
        }
      ]
      \end{axis}
    }
  }
}


\pic (beam-25) at ( 0, 0 )
  { beam_pattern_pic = {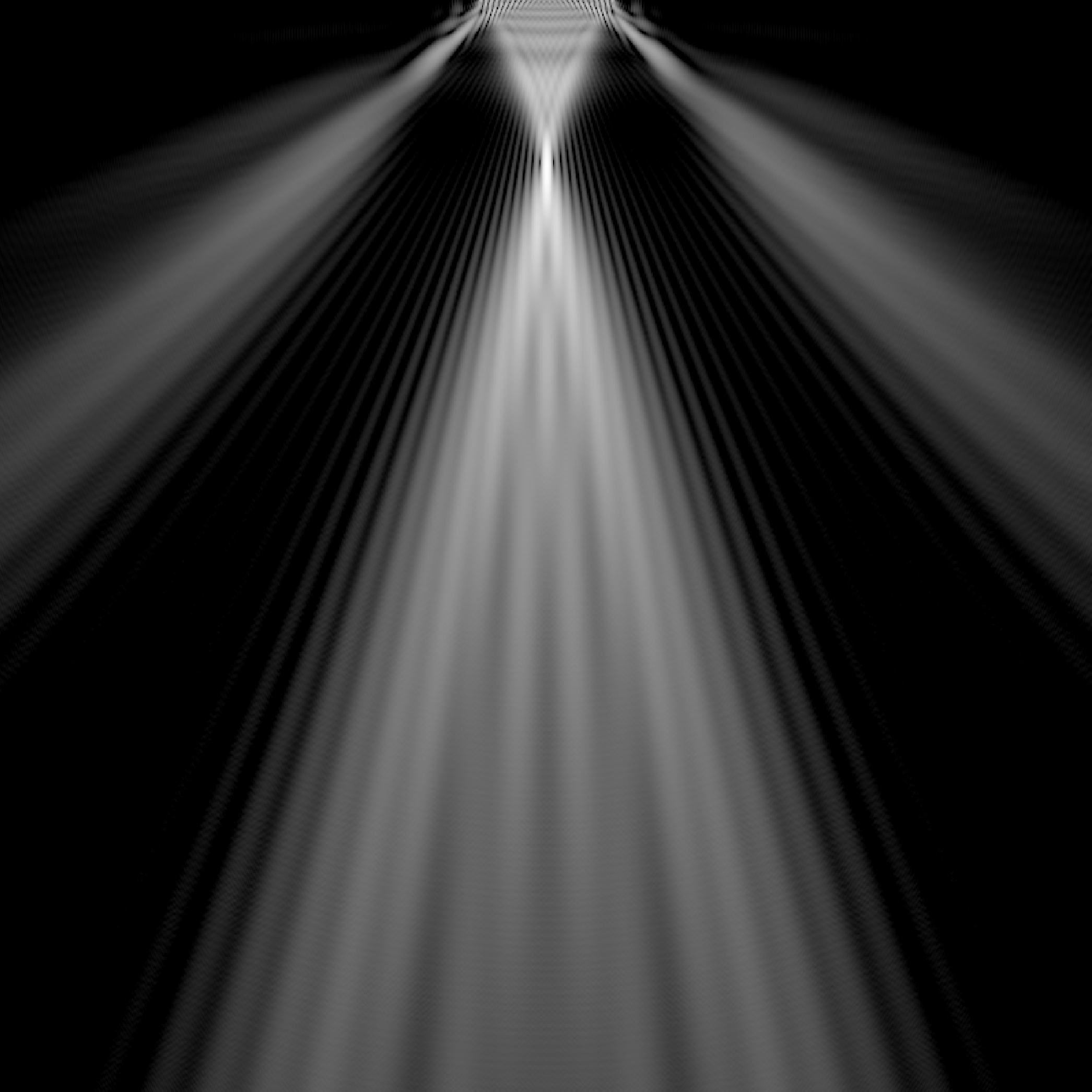}{\phantomsubcaption\label{fig:results_directivity_25}(a) \normalfont $F = 1.2$, $N_{\text{sub}} = 25$}{25}{Lateral position $x / \lambda$ (1)}{Axial position $z / \lambda$ (1)}{profile_1} };

\pic (beam-35) at ( \AxisLengthX + \AxisDistanceX, 0 )
  { beam_pattern_pic = {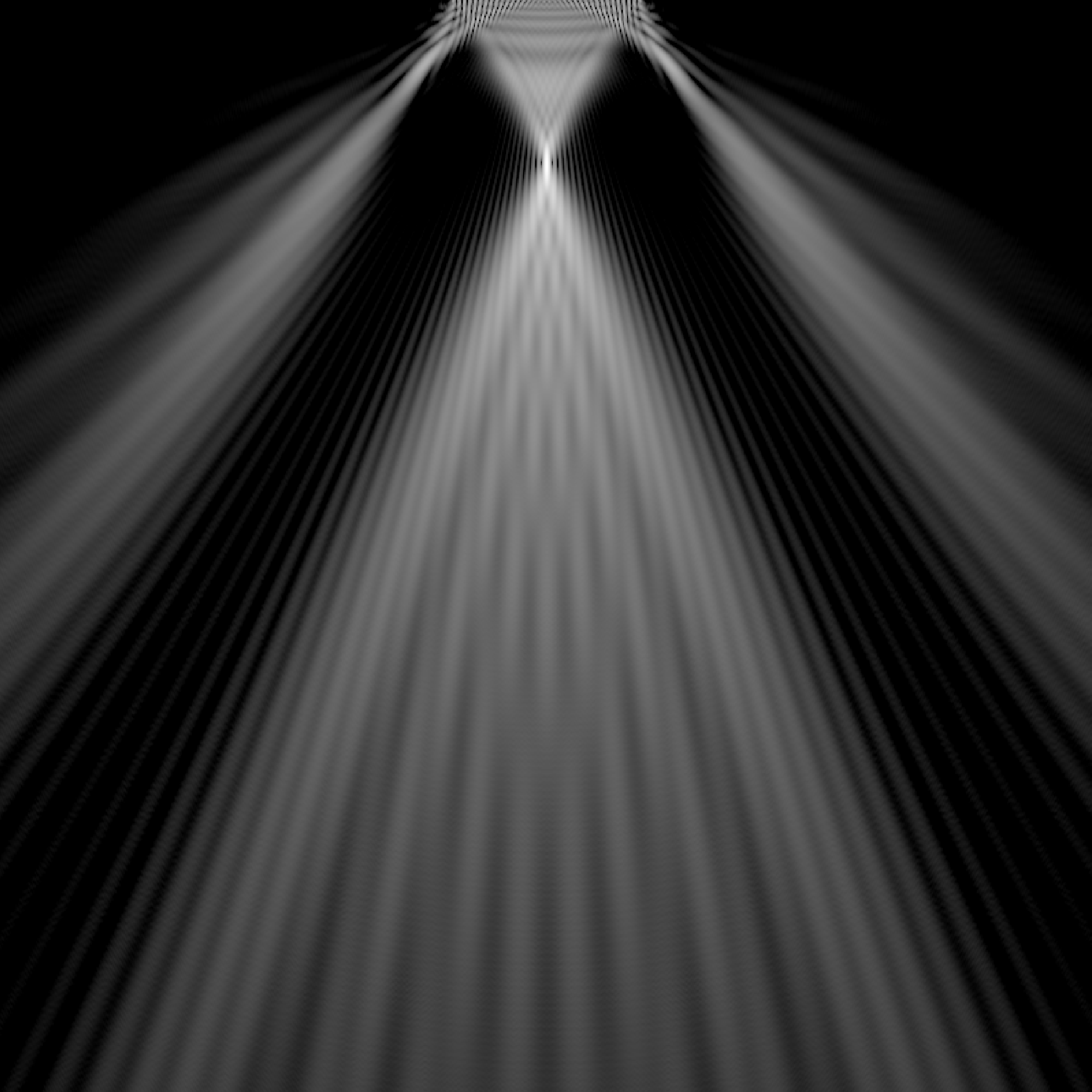}{\phantomsubcaption\label{fig:results_directivity_35}(b) \normalfont $F = 0.86$, $N_{\text{sub}} = 35$}{35}{Lateral position $x / \lambda$ (1)}{Axial position $z / \lambda$ (1)}{profile_2} };

\pic (beam-45) at ( 0, - \AxisLengthZ - \AxisDistanceZ )
  { beam_pattern_pic = {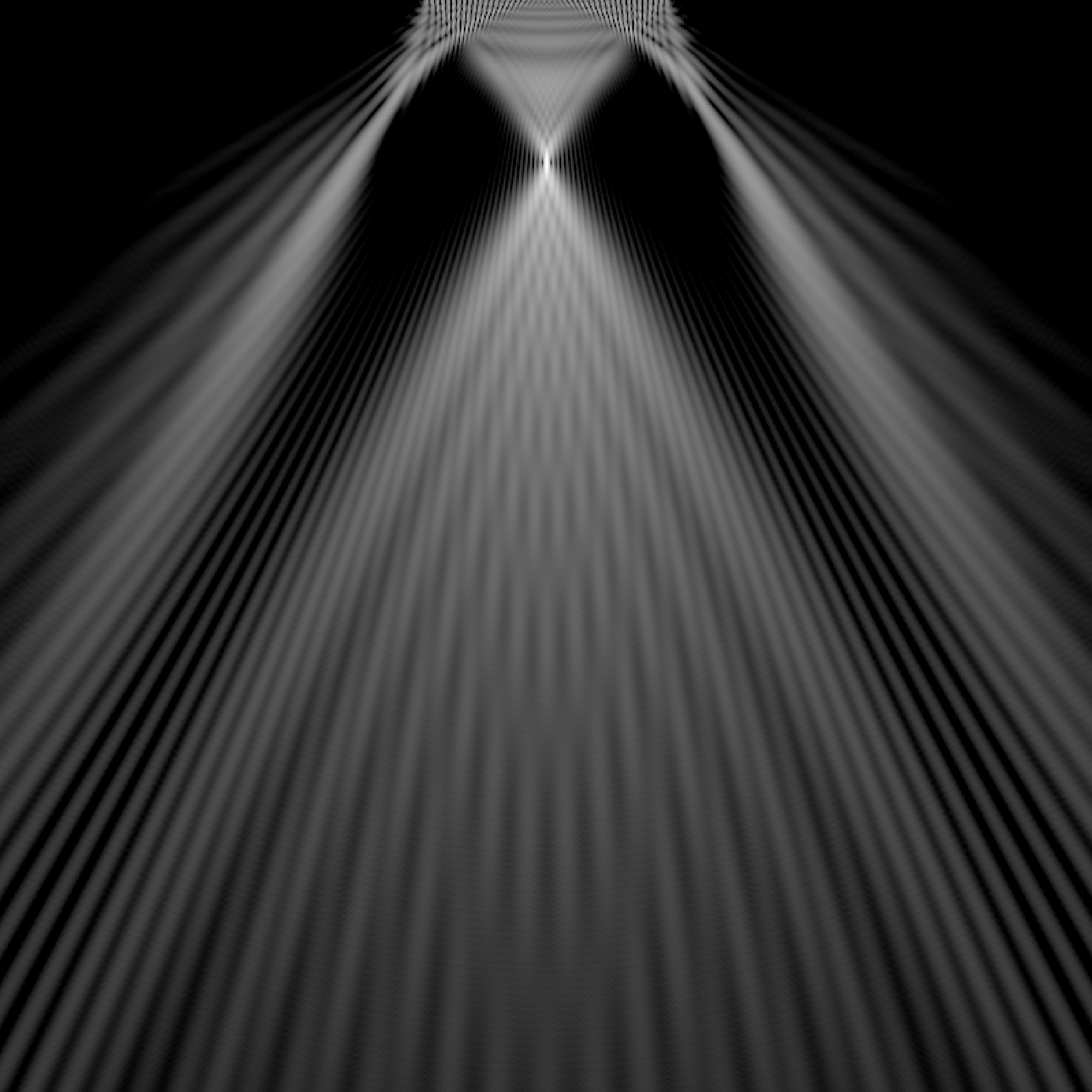}{\phantomsubcaption\label{fig:results_directivity_45}(c) \normalfont $F = 0.67$, $N_{\text{sub}} = 45$}{45}{Lateral position $x / \lambda$ (1)}{Axial position $z / \lambda$ (1)}{profile_3} };

\pic (profile) at ( \AxisLengthX + \AxisDistanceX, - \AxisLengthZ - \AxisDistanceZ )
  { profile_pic = {profile_apo_boxcar_N_elements_25_width_over_pitch_91.80.dat}{profile_apo_boxcar_N_elements_35_width_over_pitch_91.80.dat}{profile_apo_boxcar_N_elements_45_width_over_pitch_91.80.dat}{}{\phantomsubcaption\label{fig:results_directivity_profiles}(d) Profiles}{Lateral position $x / \lambda$ (1)}{Amplitude (dB)} };

\pic (nine_wires_cb) at ( 0, - 2 * \AxisLengthZ - \AxisDistanceZ - \AxisDistanceZColorbar )
  { colorbar = {-30}{0}{dB} };

\matrix [ legend, anchor = north west, minimum width = \LegendLengthX cm, nodes = {font = \footnotesize} ]
  at ( 0, - 2 * \AxisLengthZ - \AxisDistanceZ - \AxisDistanceZColorbar - 4*\ColorbarLengthY ) {%
  \node (elements_inactive) [ draw = none, fill, aperture_inactive, rounded corners = 0, minimum width = 0.5em, label = {[anchor = west, minimum width = 1em]right:Array elements} ] {}; &
  \node (elements_active) [ draw = none, fill, aperture_active, rounded corners = 0, minimum width = 0.5em, label = {[anchor = west, minimum width = 1em]right:Active array elements} ] {}; &
  \node (focus) [ focus, minimum width = 0.5em, label = {[anchor = west, minimum width = 1em]right:Focus} ] {}; \\
  \node (angular_aperture) [ minimum width = 0.5em, label = {[anchor = west, minimum width = 1em]right:Angular aperture} ] {}; &
  \node (grating_lobe) [ minimum width = 0.5em, label = {[anchor = west, minimum width = 1em]right:\phantom{First-order grating lobe}} ] {};\\
};

\draw [ focus ] (focus.west) -- (focus.east);
\draw [ focus ] (focus.north) -- (focus.south);

\draw [ alpha_lb_act, very thick ] (angular_aperture.west) -- (angular_aperture.east);

\draw [ grating_lobe, very thick ] (grating_lobe.west) -- (grating_lobe.east);
\node [ anchor = west ] at (grating_lobe.east) { First-order grating lobe bounds };

\end{tikzpicture}

%% file: theory/theory_far_field_frequency.tex
The frequency
$f$, in addition to
the $F$-number
\eqref{eqn:theory_f_number}, strongly affects
the position of
the grating lobes.
Since
there is no simple closed-form expression for
the receive beam in
the focal plane,
the far-field directivity pattern of
the focused receive subaperture will be considered here.
This pattern, for
each frequency, equals
the Fourier transform of
the aperture function and, omitting
the derivation, reveals
the identities
\begin{align}
  \sin( \alpha )
  &=
  \frac{
    1
  }{
    \sqrt{ 1 + ( 2 F )^{2} }
  }
 & \text{and} & &
  \sin( \chi )
  &=
  \frac{
    \lambda
  }{
    p
  }
  -
  \sin( \alpha ),
 \label{eqn:theory_grating_lobe_angle}
\end{align}
where
$\alpha$ denotes
the angular half-width of
the main lobe and
$\chi$ is
the angular distance of
the first-order grating lobe.
The movement of
the first-order grating lobes toward
the main lobe not only occurs for
smaller $F$-numbers
\eqref{eqn:theory_f_number} but also for
smaller wavelengths
$\lambda$ or, equivalently,
higher frequencies
$f$.

%% file: theory/theory_f_number_proposed.tex
An $F$-number that suppresses
grating lobe artifacts up to
an upper frequency bound and attempts to maintain
the lateral resolution of
the full aperture may be derived from
the angular distance of
the first-order grating lobe
\eqref{eqn:theory_grating_lobe_angle}.
The imposition of
a minimum angular distance
$\chi_{\text{lb}} \in ( 0, \pi / 2 ]$ on
this lobe yields
the closed-form expression
\begin{equation}
  F
  >
  F_{\text{lb}}( \chi_{\text{lb}}, \lambda )
  =
  \frac{
    1
  }{
    2
  }
  \sqrt{
    \frac{
      1
    }{
      \left[
        \frac{
          \lambda
        }{
          p
        }
        -
        \sin( \chi_{\text{lb}} )
      \right]^{2}
    }
    -
    1
  }
 \label{eqn:theory_f_number_proposed}
\end{equation}
for
$1 / [ 1 + \sin( \chi_{\text{lb}} ) ] < p / \lambda < 1 / \sin( \chi_{\text{lb}} )$.
This expression is
the smallest $F$-number that meets
the grating lobe condition and, thus, maximizes
the lateral resolution
(see \cref{subsec:theory_effect_receive_beam}).
Low frequencies, i.e.,
$p / \lambda \leq 1 / [ 1 + \sin( \chi_{\text{lb}} ) ]$, always permit
the usage of
the full aperture.
High frequencies, i.e.,
$p / \lambda \geq 1 / \sin( \chi_{\text{lb}} )$, however, do not support
the grating lobe condition and require
separate measures.
These measures are
outside the scope of
this paper.
Larger $F$-numbers, which may result from
the finite aperture, also meet
the grating lobe condition but unnecessarily reduce
the lateral resolution.

%% file: theory/theory_accomplishments.tex
The proposed $F$-number
\eqref{eqn:theory_f_number_proposed} not only eliminates
grating lobe artifacts but also improves
the lateral resolution.
This F-number, at high frequencies, reduces
the receive subaperture to remove
spatially undersampled components of
the \ac{RF} signals and suppress
grating lobes.
The F-number, at low frequencies, enlarges
the receive subaperture to use
the components of
all \ac{RF} signals and maintain
the lateral resolution of
the full aperture.

%% file: implementation/implementation.tex
Our Fourier-domain beamforming algorithm
\cite{proc:SchiffnerIUS2016a} was enhanced to account for
the frequency dependence of
the proposed $F$-number
\eqref{eqn:theory_f_number_proposed}.
The resulting algorithm, first, decomposes
the recorded \ac{RF} signals into
complex-valued Fourier coefficients by
the \acl{FFT}.
These coefficients, for
each frequency and
each image voxel, are then shifted in
phase to compensate for
the round-trip \aclp{TOF} and weighted by
a frequency-specific window function that reflects
the width of
the receive subaperture.
The details of
the implementation, e.g.,
the treatment of
(i) the singularity of
the proposed $F$-number
\eqref{eqn:theory_f_number_proposed} at
$\lambda / p = \sin( \chi_{\text{lb}} )$ and
(ii) edge effects near
the array bounds, are left to
an additional publication.
The first author, however, maintains
a public version of
the \name{Matlab}%
\footnote{%
  The MathWorks, Inc., Natick, MA, USA
  \label{ftnote:manufacturer_mathworks}
} source code
\cite{software:FNumber} to support
the reproduction of
the presented results and facilitate
further research.

%% file: experimental_validation/experimental_validation.tex
An experiment with
a commercial multi-tissue phantom%
\footnote{%
  Computerized Imaging Reference Systems (CIRS), Inc., Norfolk, VA, USA%
  \label{ftnote:manufacturer_cirs}
}
(%
  model: 040;
  average speed of sound: $c = \SI{1538.75}{\meter\per\second}$%
) demonstrated
the advantages of
the position- and frequency-dependent receive subaperture selection by
the proposed $F$-number
\eqref{eqn:theory_f_number_proposed}.
A SonixTouch Research system%
\footnote{%
  Analogic Corporation, Sonix Design Center, Richmond, BC, Canada
  \label{ftnote:manufacturer_analogic}
} with
a linear transducer array
(%
  model: L14-5/38;
  number of elements: $N_{\text{el}} = 128$,
  element width: $w = \SI{279.8}{\micro\meter}$,
  pitch: $p = \SI{304.8}{\micro\meter}$%
) acquired and stored
the \ac{RF} signals induced by
eleven \aclp{PW} for
offline processing.
The excitation voltage was
a single cycle at
\SI{4}{\mega\hertz}, and
the steering angles ranged from
\SI{-20}{\degree} to
\SI{20}{\degree} with
a uniform spacing of
\SI{4}{\degree}.
The lower and
upper bounds on
the frequency $f$ in
the Fourier-domain beamforming algorithm
(see \cref{sec:implementation}) were
$f_{\text{lb}} = \SI{2.25}{\mega\hertz}$ and
$f_{\text{ub}} = \SI{6.75}{\mega\hertz}$,
respectively, and resulted in
element pitch-to-wavelength ratios
$p / \lambda$ between
\num{0.45} and
\num{1.34}.
The minimum angular distance of
the first-order grating lobe in
the proposed $F$-number
\eqref{eqn:theory_f_number_proposed} was set to
$\chi_{\text{lb}} = \SI{60}{\degree}$ and permitted
the usage of
the full aperture for
element pitch-to-wavelength ratios
$p / \lambda$ up to $0.54$.
Both
the full aperture and
the smallest fixed $F$-number
\eqref{eqn:theory_f_number} that eliminated
all visible grating lobe artifacts, i.e.,
$F = 1.5$, served as
benchmarks.
The \acp{GCNR}
\cite{article:Rodriguez-MolaresITUFFC2020} of
the anechoic regions and
the axial and
lateral \acp{FWHM} of
all \num{16} wires measured
the image quality.

%% file: results/results.tex
\begin{table*}[th!]
 \centering
 \caption{%
  Properties of
  the Fixed $F$-Number
  \eqref{eqn:theory_f_number} and
  the Proposed $F$-Number
  \eqref{eqn:theory_f_number_proposed}.
 }
 \label{tab:conclusion}
 \small
 \begin{tabular}{%
  @{}%
  l
  l
  l
  l
  @{}%
 }
 \toprule
  \multicolumn{1}{@{}H}{Method} & 
  \multicolumn{1}{H}{Width of the receive aperture} &
  \multicolumn{1}{H}{Lateral resolution} &
  \multicolumn{1}{H@{}}{Grating lobe suppression}\\
  \cmidrule(r){1-1}\cmidrule(lr){2-2}\cmidrule(lr){3-3}\cmidrule(l){4-4}
 \addlinespace
  No $F$-Number & Always full & Optimal & None\\
  Fixed $F$-Number \eqref{eqn:theory_f_number} & Focal position-dependent subaperture & Minimal & Exaggerated\\
  Proposed $F$-Number \eqref{eqn:theory_f_number_proposed} & Focal position- and frequency-dependent subaperture & Improved & Optimal\\
 \addlinespace
 \bottomrule
 \end{tabular}
\end{table*}

%
\begin{figure}[t!]
 \centering%
  \input{results/figures/latex/results_experiments_cirs_040.tex}
 \caption{}
 \label{fig:V}
\end{figure}
C
{
 Results for
 the multi-tissue phantom.
 The images show
 the absolute voxel values for
 the \acl{PW} with
 the steering angle of
 \SI{-20}{\degree} and
 \subref{fig:exp_val_cirs_040_images_no}
 the full aperture,
 \subref{fig:exp_val_cirs_040_images_fixed}
 the fixed $F$-number of
 $F = 1.5$, and
 \subref{fig:exp_val_cirs_040_images_proposed}
 the proposed $F$-number
 \eqref{eqn:theory_f_number_proposed} with
 $\chi_{\text{lb}} = \SI{60}{\degree}$.
 The box plots
 \subref{fig:exp_val_cirs_040_metrics} show
 the \acp{GCNR} of
 the anechoic regions (top) and
 the lateral \acp{FWHM} of
 the wires (bottom).
}%
{exp_val_cirs_040_images}

The fixed $F$-number
\eqref{eqn:theory_f_number} eliminated
all visible grating lobe artifacts but reduced
the lateral resolution in comparison to
the full aperture, as shown in
\cref{fig:exp_val_cirs_040_images_no,fig:exp_val_cirs_040_images_fixed} for
a single plane wave.
The proposed $F$-number
\eqref{eqn:theory_f_number_proposed} also eliminated
these artifacts but increased
the lateral resolution in comparison to
the fixed $F$-number, as shown in
\cref{fig:exp_val_cirs_040_images_proposed}.
%
The median anechoic contrast, according to
\cref{fig:exp_val_cirs_040_metrics}, amounted to
\SI{87.38}{\percent} for
eleven steering angles and, thus, improved by
\SI{2}{\percent} and
\SI{3.2}{\percent} relative to
the full aperture and
the fixed $F$-number,
respectively.
%
The median \acp{FWHM} of
the wires on
the axial and
lateral axes decreased by
up to \SI{2.9}{\percent} and
\SI{12.8}{\percent},
respectively, resulting in
a volume reduction of
up to \SI{19.6}{\percent} for
a single plane wave.

%% file: results/figures/latex/results_experiments_cirs_040.tex
%
\begin{tikzpicture}
[
  font = \footnotesize,
  arrow_indicator/.style = { white, thick, ->, >=Stealth },
  title/.style = { inner sep = 0, outer sep = 0, anchor = south west, font = \bfseries },
  f_number/.style = { white, anchor = south west },
  no_f_number/.style = { draw = none, fill = RUBGRAY_RGB, rounded corners = 0, minimum width = 0.5em },
  fixed_f_number/.style = { draw = none, fill = RUBBLUE_RGB, rounded corners = 0, minimum width = 0.5em },
  proposed_f_number/.style = { draw = none, fill = RUBGREEN_RGB, rounded corners = 0, minimum width = 0.5em },
  legend/.style = { draw = black!50, fill = black!5, rounded corners = 5, thick, outer sep = 0 }%
]

\pgfmathsetmacro{\AxisLengthX}{3.4}   
\pgfmathsetmacro{\AxisLengthZ}{3.4}   

\pgfmathsetmacro{\AxisMetricsLengthX}{3.4}   
\pgfmathsetmacro{\AxisMetricsLengthZNorm}{0.4}

\pgfmathsetmacro{\AxisDistanceX}{1.1}  
\pgfmathsetmacro{\AxisDistanceZ}{1.5}  
\pgfmathsetmacro{\AxisDistanceZColorbar}{1}  

\pgfmathsetmacro{\ColorbarLengthY}{0.2}     

\pgfmathsetmacro{\ImageLimitsXLBCIRS}{-19.5072} 
\pgfmathsetmacro{\ImageLimitsXUBCIRS}{19.5072}  
\pgfmathsetmacro{\ImageLimitsZLBCIRS}{4.8768}   
\pgfmathsetmacro{\ImageLimitsZUBCIRS}{43.8912}  

\pgfmathsetmacro{\AxisLimitsXLBCIRS}{-19.25}  
\pgfmathsetmacro{\AxisLimitsXUBCIRS}{19.25}   
\pgfmathsetmacro{\AxisLimitsZLBCIRS}{5.4}     
\pgfmathsetmacro{\AxisLimitsZUBCIRS}{43.9}    

\pgfmathsetmacro{\MagCtrX}{1.92625}         
\pgfmathsetmacro{\MagCtrZ}{25.9644}         
\pgfmathsetmacro{\MagLengthX}{3}            
\pgfmathsetmacro{\MagLengthZ}{3}            
         
\pgfmathsetmacro{\MagImageSizeRel}{0.3}     
\pgfmathsetmacro{\MagImageShiftX}{0.01}     
\pgfmathsetmacro{\MagImageShiftZ}{0.01}     

\pgfmathsetmacro{\MagScaleLengthX}{1}       
\pgfmathsetmacro{\MagScaleLengthZ}{2}       
\pgfmathsetmacro{\MagScalePosRelZ}{0.35}    

\pgfmathsetmacro{\AxisLabelXShift}{-0.15}  
\pgfmathsetmacro{\AxisLabelZShift}{-0.18}  

\pgfmathsetmacro{\ArrowLength}{5}
\pgfmathsetmacro{\ArrowAngle}{125}

\pgfmathsetmacro{\PositionArrowX}{-5}
\pgfmathsetmacro{\PositionArrowZ}{22}

\pgfmathsetmacro{\MagLimitsXLB}{ \MagCtrX - 0.5 * \MagLengthX } 
\pgfmathsetmacro{\MagLimitsXUB}{ \MagCtrX + 0.5 * \MagLengthX } 
\pgfmathsetmacro{\MagLimitsZLB}{ \MagCtrZ - 0.5 * \MagLengthZ } 
\pgfmathsetmacro{\MagLimitsZUB}{ \MagCtrZ + 0.5 * \MagLengthZ } 
\pgfmathsetmacro{\MagImageSizeX}{ \MagImageSizeRel * \AxisLengthX }
\pgfmathsetmacro{\MagImageSizeZ}{ \MagImageSizeRel * \AxisLengthZ }
\pgfmathsetmacro{\MagImagePosX}{ \MagImageShiftX * \AxisLengthX }
\pgfmathsetmacro{\MagImagePosZ}{ \MagImageShiftZ * \AxisLengthZ }
\pgfmathsetmacro{\MagScaleXLB}{ \MagCtrX - 0.5 * \MagScaleLengthX } 
\pgfmathsetmacro{\MagScaleXUB}{ \MagCtrX + 0.5 * \MagScaleLengthX } 
\pgfmathsetmacro{\MagScaleZ}{ \MagCtrZ + \MagScalePosRelZ * 0.5 * \MagLengthZ }
\pgfmathsetmacro{\MagScaleLengthZImage}{ \MagScaleLengthZ * ( \MagLimitsZUB - \MagLimitsZLB ) / ( 10 * \MagImageSizeRel * \MagLengthZ ) }
\pgfmathsetmacro{\MagScaleZLB}{ \MagScaleZ - 0.5 * \MagScaleLengthZImage }
\pgfmathsetmacro{\MagScaleZUB}{ \MagScaleZ + 0.5 * \MagScaleLengthZImage }

\pgfmathsetmacro{\AxisMetricsLengthZ}{ \AxisMetricsLengthZNorm * \AxisLengthZ }   

\pgfmathsetmacro{\ColorbarLengthX}{ 2 * \AxisLengthX + \AxisDistanceX }

\pgfmathsetmacro{\LegendLengthX}{ \ColorbarLengthX }

\tikzset{
  pics/cirs_040_phantom/.style n args = {5}{%
    code = {%
      \begin{axis}
      [%
        width = \AxisLengthX cm, height = \AxisLengthZ cm,
        scale only axis = true,
        at = { ( 0, 0 ) },
        xmin = \AxisLimitsXLBCIRS, xmax = \AxisLimitsXUBCIRS,
        ymin = \AxisLimitsZLBCIRS, ymax = \AxisLimitsZUBCIRS,
        y dir = reverse,
        axis x line = bottom,
        axis y line = left,
        axis line style = { - },
        every axis title/.append style = { title, at = { ( -0.15, 1 ) } },
        title = { #2 },
        xtick = { -20, -15, -10, -5, 0, 5, 10, 15, 20 },
        xticklabel style = { font = \scriptsize },
        ytick = { 5, 10, 15, 20, 25, 30, 35, 40, 45},
        yticklabel style = { font = \scriptsize },
        x label style = { at = { (axis description cs:0.5,\AxisLabelXShift) }, anchor = north, outer sep = 0, inner sep = 0 },
        y label style = { at = { (axis description cs:\AxisLabelZShift,0.5) }, anchor = south, outer sep = 0, inner sep = 0 },
        xlabel = { #4 },
        ylabel = { #5 },
        name = -axis
      ]

        \addplot [ forget plot ] graphics [ xmin = \ImageLimitsXLBCIRS, xmax = \ImageLimitsXUBCIRS, ymin = \ImageLimitsZLBCIRS, ymax = \ImageLimitsZUBCIRS ]
          { #1 };

        \pgfmathsetmacro{\StopX}{ \PositionArrowX + cos(\ArrowAngle) * \ArrowLength }
        \pgfmathsetmacro{\StopZ}{ \PositionArrowZ - sin(\ArrowAngle) * \ArrowLength }
        \draw [ arrow_indicator ] (\PositionArrowX,\PositionArrowZ) -- (\StopX,\StopZ);

        \draw [ draw = white ] (\MagLimitsXLB, \MagLimitsZLB) rectangle (\MagLimitsXUB, \MagLimitsZUB);

        \node [ f_number ] at (\AxisLimitsXLBCIRS, \AxisLimitsZUBCIRS) { #3 };

      \end{axis}
    }
  }
}

\tikzset{
  pics/cirs_040_phantom_mag/.style n args = {2}{%
    code = {%
      \begin{axis}
      [%
         width = \MagImageSizeX cm, height = \MagImageSizeZ cm,
         scale only axis = true,
         at = { ( 0.1, 0.1 ) },
         xmin = \MagLimitsXLB, xmax = \MagLimitsXUB,
         ymin = \MagLimitsZLB, ymax = \MagLimitsZUB,
         y dir = reverse,
         axis lines = box,
         axis line style = { -, white, thick },
          tick style = { draw = none },
          xtick = {\empty},
          ytick = {\empty},
        x label style = { at = { (axis description cs:0.5,\AxisLabelXShift) }, anchor = north, outer sep = 0, inner sep = 0 },
        y label style = { at = { (axis description cs:\AxisLabelZShift,0.5) }, anchor = south, outer sep = 0, inner sep = 0 },
        xlabel = { },
        ylabel = { },
        name = -axis-mag
      ]

        \addplot [ forget plot ] graphics [ xmin = \ImageLimitsXLBCIRS, xmax = \ImageLimitsXUBCIRS, ymin = \ImageLimitsZLBCIRS, ymax = \ImageLimitsZUBCIRS ]
          { #1 };

        \draw [ white, thick ] (\MagScaleXLB, \MagScaleZ) -- (\MagScaleXUB, \MagScaleZ)
         node [ pos = 0.5, font = \scriptsize, anchor = north, outer sep = 0 ] { \SI{\MagScaleLengthX}{\milli\meter} };
        \draw [ white, thick ] (\MagScaleXLB, \MagScaleZLB) -- (\MagScaleXLB, \MagScaleZUB);
        \draw [ white, thick ] (\MagScaleXUB, \MagScaleZLB) -- (\MagScaleXUB, \MagScaleZUB);

      \end{axis}
    }
  }
}

\tikzset{
  pics/colorbar/.style n args = {3}{%
    code = {%
      \begin{axis}
      [%
         width = 0, height = 0,
         scale only axis = true,
         at = { ( 0, 0 ) },
         hide axis,
         xtick = {\empty}, ytick = {\empty},
         point meta min = #1, point meta max = #2,
         colormap/blackwhite,
         colorbar horizontal,
         colorbar style = {
           at = { ( 0, 0 ) },
           xtick = { #1, -50, -40, -30, -20, -10, #2 },
           xticklabel style = { font = \scriptsize },
           anchor = north west,
           width = \ColorbarLengthX cm, height = \ColorbarLengthY cm,
           y label style = { at = { (axis description cs:0,0.5) }, anchor = east, rotate = -90 },
           ylabel = { #3 }
         }
      ]
      \end{axis}
    }
  }
}

\tikzset{
  pics/box_plot_fwhm/.style n args = {4}{%
    code = {%
      \begin{axis}
      [%
         width = \AxisMetricsLengthX cm, height = \AxisMetricsLengthZ cm,
         scale only axis = true,
         at = { ( 0, 0 ) },
         boxplot/draw direction = y,
         ymin = 0, ymax = 1.5,
         axis x line = bottom,
         axis y line = left,
         axis line style = { - },
         every axis title/.append style = { title, at = { ( -0.15, 1 ) } },
         title = { #2 },
         xtick = { 2, 6, 10 },
         xticklabel style = { font = \scriptsize },
         xticklabels = { 1, 3, 11 },
         ytick = { 0, 0.25, 0.5, 0.75, 1, 1.25, 1.5 },
         yticklabel style = { font = \scriptsize },
         yticklabels = { 0, \empty, 0.5, \empty, 1, \empty, 1.5 },
         grid = major,
        x label style = { at = { (axis description cs:0.5,\AxisLabelXShift / \AxisMetricsLengthZNorm) }, anchor = north, outer sep = 0, inner sep = 0 },
        y label style = { at = { (axis description cs:\AxisLabelZShift,0.5) }, anchor = south, outer sep = 0, inner sep = 0 },
        xlabel = { #3 },
        ylabel = { #4 },
        name = -axis
      ]
        \addplot [ no_f_number, boxplot prepared = { lower whisker = 0.3238, lower quartile = 0.4324, median = 0.4648, upper quartile = 0.5258, upper whisker = 0.6820, draw position = 1 } ] coordinates {};
        \addplot [ fixed_f_number, boxplot prepared = { lower whisker = 0.6934, lower quartile = 0.7239, median = 0.7429, upper quartile = 0.9335, upper whisker = 1.1963, draw position = 2 } ] coordinates {};
        \addplot [ proposed_f_number, boxplot prepared = { lower whisker = 0.5867, lower quartile = 0.6267, median = 0.6477, upper quartile = 0.8153, upper whisker = 0.9639, draw position = 3 } ] coordinates {};

        \addplot [ no_f_number, boxplot prepared = { lower whisker = 0.2476, lower quartile = 0.3391, median = 0.3562, upper quartile = 0.4572, upper whisker = 0.5715, draw position = 5 } ] coordinates {};
        \addplot [ fixed_f_number, boxplot prepared = { lower whisker = 0.4534, lower quartile = 0.4763, median = 0.4934, upper quartile = 0.6820, upper whisker = 0.8382, draw position = 6 } ] coordinates {};
        \addplot [ proposed_f_number, boxplot prepared = { lower whisker = 0.4039, lower quartile = 0.4191, median = 0.4381, upper quartile = 0.6325, upper whisker = 0.7277, draw position = 7 } ] coordinates {};

        \addplot [ no_f_number, boxplot prepared = { lower whisker = 0.2515, lower quartile = 0.3391, median = 0.3505, upper quartile = 0.4286, upper whisker = 0.5524, draw position = 9 } ] coordinates {};
        \addplot [ fixed_f_number, boxplot prepared = { lower whisker = 0.4381, lower quartile = 0.4610, median = 0.4801, upper quartile = 0.6134, upper whisker = 0.7811, draw position = 10 } ] coordinates {};
        \addplot [ proposed_f_number, boxplot prepared = { lower whisker = 0.3924, lower quartile = 0.4077, median = 0.4286, upper quartile = 0.5658, upper whisker = 0.7010, draw position = 11 } ] coordinates {};

      \end{axis}
    }
  }
}



\tikzset{
  pics/box_plot_gcnr/.style n args = {4}{%
    code = {%
      \begin{axis}
      [%
         width = \AxisMetricsLengthX cm, height = \AxisMetricsLengthZ cm,
         scale only axis = true,
         at = { ( 0, 0 ) },
         boxplot/draw direction = y,
         ymin = 0, ymax = 100,
         axis x line = bottom,
         axis y line = left,
         axis line style = { - },
         every axis title/.append style = { title, at = { ( -0.15, 1 ) } },
         title = { #2 },
         xtick = { 2, 6, 10 },
         xticklabel style = { font = \scriptsize },
         xticklabels = { 1, 3, 11 },
         ytick = { 0, 20, 40, 60, 80, 100 },
         yticklabel style = { font = \scriptsize },
         yticklabels = { 0, 20, 40, 60, 80, 100 },
         grid = major,
        x label style = { at = { (axis description cs:0.5,\AxisLabelXShift / \AxisMetricsLengthZNorm) }, anchor = north, outer sep = 0, inner sep = 0 },
        y label style = { at = { (axis description cs:\AxisLabelZShift,0.5) }, anchor = south, outer sep = 0, inner sep = 0 },
        xlabel = { #3 },
        ylabel = { #4 },
        name = -axis
      ]
        \addplot [ no_f_number, boxplot prepared = { lower whisker = 48.89, lower quartile = 48.89, median = 57.94, upper quartile = 66.98, upper whisker = 66.98, draw position = 1 } ] coordinates {};
        \addplot [ fixed_f_number, boxplot prepared = { lower whisker = 50.40, lower quartile = 50.40, median = 58.01, upper quartile = 65.62, upper whisker = 65.62, draw position = 2 } ] coordinates {};
        \addplot [ proposed_f_number, boxplot prepared = { lower whisker = 54.52, lower quartile = 54.52, median = 57.98, upper quartile = 61.45, upper whisker = 61.45, draw position = 3 } ] coordinates {};

        \addplot [ no_f_number, boxplot prepared = { lower whisker = 52.52, lower quartile = 52.52, median = 61.17, upper quartile = 69.81, upper whisker = 69.81, draw position = 5 } ] coordinates {};
        \addplot [ fixed_f_number, boxplot prepared = { lower whisker = 59.13, lower quartile = 59.13, median = 66.71, upper quartile = 74.29, upper whisker = 74.29, draw position = 6 } ] coordinates {};
        \addplot [ proposed_f_number, boxplot prepared = { lower whisker = 60.74, lower quartile = 60.74, median = 66.63, upper quartile = 72.52, upper whisker = 72.52, draw position = 7 } ] coordinates {};

        \addplot [ no_f_number, boxplot prepared = { lower whisker = 79.54, lower quartile = 79.54, median = 85.36, upper quartile = 91.19, upper whisker = 91.19, draw position = 9 } ] coordinates {};
        \addplot [ fixed_f_number, boxplot prepared = { lower whisker = 81.84, lower quartile = 81.84, median = 84.17, upper quartile = 86.49, upper whisker = 86.49, draw position = 10 } ] coordinates {};
        \addplot [ proposed_f_number, boxplot prepared = { lower whisker = 84.28, lower quartile = 84.28, median = 87.38, upper quartile = 90.47, upper whisker = 90.47, draw position = 11 } ] coordinates {};

      \end{axis}
    }
  }
}




\pic (cirs_040_0) at ( 0, 0 )
  { cirs_040_phantom = {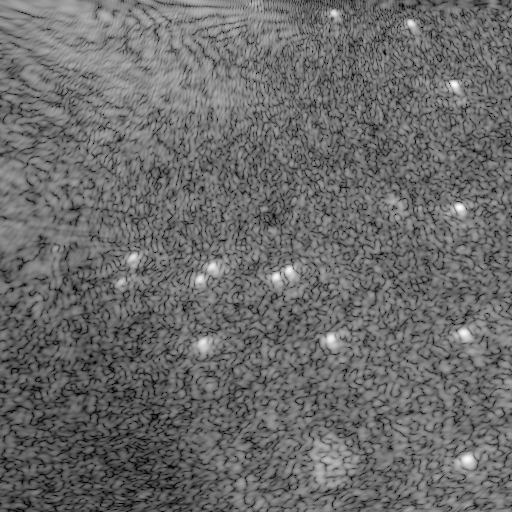}{\phantomsubcaption\label{fig:exp_val_cirs_040_images_no}(a) No $F$-number}{}{Lateral position $x$ (mm)}{Axial position $z$ (mm)} };

\pic (cirs_040_1) at ( \AxisLengthX + \AxisDistanceX, 0 )
  { cirs_040_phantom = {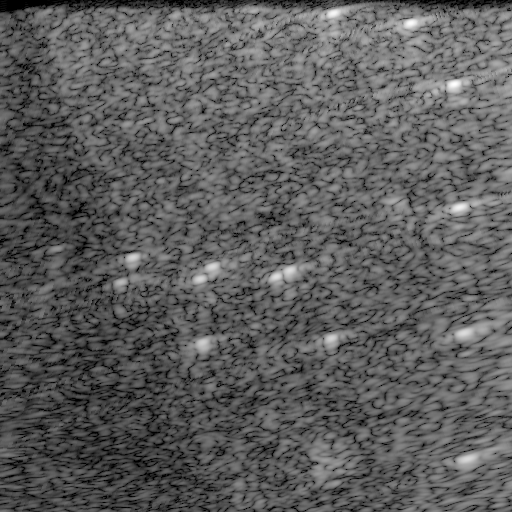}{\phantomsubcaption\label{fig:exp_val_cirs_040_images_fixed}(b) Fixed $F$-number}{}{Lateral position $x$ (mm)}{Axial position $z$ (mm)} };

\pic (cirs_040_4) at ( 0, -\AxisLengthZ - \AxisDistanceZ )
  { cirs_040_phantom = {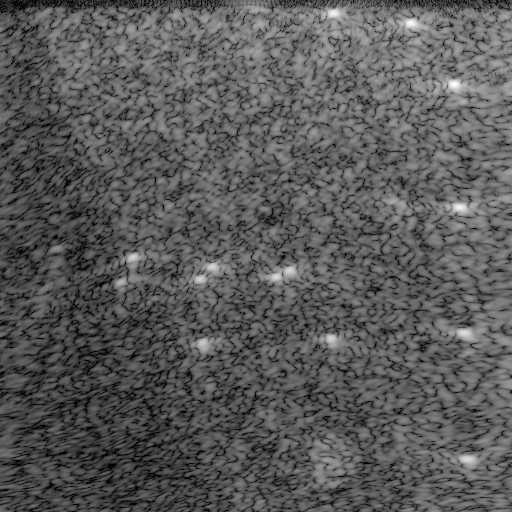}{\phantomsubcaption\label{fig:exp_val_cirs_040_images_proposed}(c) Proposed $F$-number}{}{Lateral position $x$ (mm)}{Axial position $z$ (mm)} };

\pic (cirs_040_mag_0) at ( \MagImagePosX, \MagImagePosZ )
  { cirs_040_phantom_mag = {results/figures/exp_val_cirs_040_05dB_images_constant_F_0.00_dyn_range_60.png}{} };

\pic (cirs_040_mag_1) at ( \AxisLengthX + \AxisDistanceX + \MagImagePosX, \MagImagePosZ )
  { cirs_040_phantom_mag = {results/figures/exp_val_cirs_040_05dB_images_constant_F_1.50_dyn_range_60.png}{} };

\pic (cirs_040_mag_4) at ( \MagImagePosX, -\AxisLengthZ - \AxisDistanceZ + \MagImagePosZ )
  { cirs_040_phantom_mag = {results/figures/exp_val_cirs_040_05dB_images_angle_lb_deg_60.00_F_ub_1.50_dyn_range_60.png}{} };

\pic (metrics_FWHM_contrast) at ( \AxisLengthX + \AxisDistanceX, -\AxisMetricsLengthZ - \AxisDistanceZ )
  { box_plot_gcnr = {}{\phantomsubcaption\label{fig:exp_val_cirs_040_metrics}(d) Quality metrics}{}{gCNR (\si{\percent})} };

\pic (metrics_FWHM_lateral) at ( \AxisLengthX + \AxisDistanceX, -\AxisLengthZ - \AxisDistanceZ )
  { box_plot_fwhm = {}{}{Number of angles (1)}{Lateral FWHM (mm)} };

\pic at ( 0, -\AxisLengthZ - \AxisDistanceZ - \AxisDistanceZColorbar )
  { colorbar = {-60}{0}{dB} };

\matrix [ legend, anchor = north west, minimum width = \LegendLengthX cm, nodes = {font = \footnotesize} ]
  at ( 0, - \AxisLengthZ - \AxisDistanceZ - \AxisDistanceZColorbar - 4*\ColorbarLengthY ) {%
  \node (no_F_number) [ no_f_number, label = {[anchor = west, minimum width = 1em]right:No $F$-number} ] {}; &
  \node (fixed_F_number) [ fixed_f_number, label = {[anchor = west, minimum width = 1em]right:Fixed $F$-number} ] {}; &
  \node (proposed_F_number) [ proposed_f_number, label = {[anchor = west, minimum width = 1em]right:Proposed $F$-number} ] {}; \\
};

\end{tikzpicture}%

%% file: conclusion/conclusion.tex
The proposed combination of
the frequency-dependent $F$-number
\eqref{eqn:theory_f_number_proposed} and
the Fourier-domain beamforming algorithm
(see \cref{sec:implementation}) improves
the contrast and
the lateral resolution, as summarized in
\cref{tab:conclusion}.
Details of
the implementation, such as
the treatments of
(i) the singularity of
the proposed $F$-number
\eqref{eqn:theory_f_number_proposed} at
$\lambda / p = \sin( \chi_{\text{lb}} )$ and
(ii) edge effects near
the array bounds, strongly influence
the results and enable
further improvements.
The proposed combination is
easier to implement and
potentially faster than
adaptive methods, such as
\acl{MV} beamforming
\cite{article:SynnevagITUFFC2009}.
The frequency-dependent apodization weights, in fact, are independent of
the \ac{RF} signals and may be precomputed for
a given geometry and
bandwidth.
Future research will investigate
an efficient translation of
the proposed combination into
the time domain.
Such a translation could base on
filter banks and
the usage of
a piecewise constant $F$-number.
This $F$-number would be fixed in
each frequency band to enable
standard time-domain receive beamforming.
The $F$-number, however, would vary with
the frequency band.